\documentclass[aps, prb,twocolumn,showpacs,preprintnumbers,amsmath,amssymb]{revtex4-1}
\usepackage{amssymb}
\usepackage{graphicx}
\usepackage{dcolumn}
\usepackage{bm}
\usepackage{subfigure}

\begin{document}
\preprint{PRB/Y. K. Li et al.}

\title{Magnetic phase diagram in the Co-rich side of LnCo$_{1-x}$Fe$_{x}$AsO (Ln=La, Sm) system}
\author{Y. K. Li$^{1,2}$\footnote[1]{Electronic address: yklee@hznu.edu.cn}, X. F. Xu$^{1}$, C. Cao$^{1}$, C. Y. Shen$^{2}$, Y. K. Luo$^{2}$, Q. Tao$^{2}$, X. Lin$^{1}$, L. Zhang$^{3}$, G. H. Cao$^{2}$ and Z. A. Xu$^{2}$}

\affiliation{$^{1}$Department of Physics, Hangzhou Normal University, Hangzhou 310036, China\\
$^{2}$State Key Lab of Silicon Materials and Department of Physics, Zhejiang University, Hangzhou 310027, China\\
$^{3}$Department of Physics, China Jiliang University, Hangzhou 310018, China}

\date{\today}

\begin{abstract}

The magnetic phase diagram has been mapped out via the measurements
of electronic resistivity, magnetization and specific heat in the
cobalt-based layered \emph{Ln}Co$_{1-x}$Fe$_{x}$AsO (\emph{Ln}$=$La, Sm) compounds.
The ferromagnetic (FM) transition at $\sim$63 K for LaCoAsO is rapidly
suppressed upon Fe doping, and ultimately disappears around $x$ $=$
0.3 in the LaCo$_{1-x}$Fe$_{x}$AsO system. When La is replaced by
magnetic rare earth element Sm, the $3d$--electrons first undergo a
FM transition at \emph{T$_{c}$}$\sim 75$ K, followed by an antiferromagnetic (AFM) transition at a lower temperature \emph{T$_{N1}$}$\sim 45$ K. With partial Fe doping on the Co site, both FM (\emph{T$_{c}$}) and AFM (\emph{T$_{N1}$}) transition temperatures are significantly suppressed, and finally approach zero kelvin at $x$ $=$ 0.3 and 0.2, respectively.
Meanwhile, a third magnetic transition at \emph{$T_{N2}$}$\sim 5.6$ K for
SmCoAsO, associated with the AFM order of the Sm$^{3+}$
4f--moments, is uncovered and \emph{$T_{N2}$} is found to be almost
robust against the small Fe-doping. These results suggest that the 4f--electrons of Sm$^{3+}$ have an important effect
on the magnetic behavior of 3d electrons in the 1111 type Co-based
\emph{Ln}Co$_{1-x}$Fe$_{x}$AsO systems. In contrast, the magnetism of the
f--electrons is relatively unaffected by the variation of the 3d
electrons. The rich magnetic phase diagram in the Co-rich side of
the \emph{Ln}Co$_{1-x}$Fe$_{x}$AsO system, therefore, is established.

\end{abstract}
\maketitle

\section{\label{sec:level1}Introduction}

Correlated electron systems \emph{LnTmPn}O [\emph{Ln}$=$rare earth element, \emph{Tm}$=$transition metal element, \emph{Pn}$=$pnictogen element] have attracted great attention due to their various electronic and magnetic properties,
such as high transition temperature superconductivity\cite{Hosono}, itinerant ferromagnetism\cite{LaCoAsO,Yoshimura2},
giant magnetoresistance\cite{Yoshimura}, spin density wave (SDW)\cite{Dai} and structural instability\cite{Structure}.
For example, iso-structural LaOMnAs is an AFM semiconductor\cite{LaMnAsO}, and LaONiAs shows
superconductivity below 3 K\cite{LuoNi}. In the case of \emph{Tm}$=$Co, LnCoAsO was
reported to be an itinerant ferromagnet with the Curie temperature \emph{T$_{c}$} between 60 and 80
K for La and Sm, respectively\cite{LaCoAsO,LnCoAsO}. Among them, the compound LaFeAsO, which is a parent compound of well-known iron-based superconductors, exhibits a spin-density wave antiferromagnetic transition at about 150 K\cite{Hosono}. When Fe is partially replaced by Co atoms, the AFM order from Fe ions is suppressed and then superconductivity emerges, and the compound exhibits a good metallic behavior down
to superconducting (SC) transition temperature\cite{SefatCo,WangC}. Similar results have also been reported for Co-doped CeFeAsO\cite{CeCo}, PrFeAsO\cite{PrCo,PrCo2}, NdFeAsO\cite{NdCoFe} and SmFeAsO\cite{WangC} systems. Thus, magnetism is
closely related to superconductivity in these iron-based high-temperature superconductors.

On the other hand, the \emph{Ln}CoAsO compounds (also referred to as Co-1111 system ) with the same space
group as ZrCuSiAs exhibit rich magnetic properties at low temperature. LaCoAsO is reported
to be an itinerant ferromagnet with 2D ferromagnetic spin fluctuations\cite{LaCoAsO,NMR}. When La
is substituted by other magnetic rare earth elements, \emph{Ln}CoAsO (\emph{Ln}$=$Nd, Sm, and
Gd)\cite{Mcgurie,SmCoAsO,SmCoAsO2,LiCo,SmuSR,LnCoAsO} undergoes multiple magnetic phase transitions as the
temperature decreases. Furthermore, the AFM order due to the magnetic sublattice of \emph{Ln} ions at very
low temperature can be also observed in those compounds, almost irrelevant to the doping at \emph{Tm} or \emph{Pn} sites.
For example, in the case of SmCoAsO\cite{LiCo}, a ferromagnetic (FM) transition occurs around
\emph{T$_{c}$ }of 75 K, followed by a FM-AFM transition from the magnetic coupling between the CoAs layers
around 45 K, and finally another AFM order from Sm ion forms at 5.6 K recently reported by other groups\cite{SmCoAsO,Sm-AFM,Sm-AFM2}. Indeed, several groups\cite{LnCoAsO,RKKY} have suggested that the Ruderman-Kittel-Kasuya-Yoshida (RKKY) interaction may play a role in the FM-AFM transition. It is ascribed to the interaction between the localized magnetic moments of lanthanide 4f electrons and
the ferromagnetic ordered magnetic moments of cobalt 3d itinerant electrons. The neutron diffraction experiments\cite{Mcgurie,NdCoFe} and specific heat measurements\cite{SmCoAsO,Sm-AFM,Sm-AFM2} have detected the localized magnetic moment of \emph{Ln} 4f electrons in those parent compounds.

Up to now, main studies about Co-containing 1111 system focus on these low Co concentrations\cite{SefatCo,WangC,CeCo,PrCo,PrCo2,NdCoFe} and
\emph{Ln}Co\emph{Pn}O parent compounds\cite{LaCoAsO,Yoshimura,LnCoAsO,Mcgurie,SmCoAsO,SmCoAsO2,LiCo}. There are few reports on the study of chemical doping in \emph{Ln}CoAsO\cite{NdFeCo}, and the magnetic phase diagram on the Co-rich side of \emph{Ln}CoAsO is less known. In this paper, we report our detailed study of the magnetic properties of Fe-doped \emph{Ln}Co$_{1-x}$Fe$_{x}$AsO (\emph{Ln}$=$La, Sm)
system on the Co-rich side. In order to study the interplay between 4f electrons and 3d
electrons, LaCo$_{1-x}$Fe$_{x}$AsO system is employed as a comparison. We performed powder X-ray
diffraction, electrical resistivity, and magnetization measurements, as well as the
first-principles calculations. The results of these measurements and calculations
indicate that, the FM order is quickly suppressed by Fe doping in LaCo$_{1-x}$Fe$_{x}$AsO system,
and finally disappears at about $x$ $=$ 0.3. In the case of SmCo$_{1-x}$Fe$_{x}$AsO, the FM and AFM
transitions of the 3d--electrons are gradually suppressed and then disappear at $x$ $=$ 0.3 and
0.2, respectively. However, the AFM order at low temperature due to Sm$^{3+}$ is robust and
\emph{T$_{N2}$} slightly varies with increasing Fe content. A rich magnetic phase diagram for
$x$ $\leq$ 0.3 \emph{Ln}Co$_{1-x}$Fe$_{x}$AsO system is therefore established.

\section{\label{sec:level1}Experimental}

The polycrystalline samples of \emph{Ln}Co$_{1-x}$Fe$_{x}$AsO (\emph{Ln}$=$La, Sm ) were
synthesized by two-step solid state reaction methods in vacuum,
similar to our previous reports\cite{WangC}. The pellets of \emph{Ln}Co$_{1-x}$Fe$_{x}$AsO ($x$ $=$ 0, 0.05, 0.1, 0.2, 0.3,) were annealed in an
evacuated quartz tube at 1423 K for 40 hours and furnace-cooled to
room temperature.

Crystal structure measurement was performed by powder X-ray
diffraction (XRD) at room temperature using a D/Max-rA
diffractometer with Cu K$_{\alpha}$ radiation and a graphite
monochromator. Lattice parameters were calculated by least-squares
fitting using at least 20 XRD peaks. The electrical resistivity was
measured by four-terminal method. The temperature dependence of d.c.
magnetization was measured on a Quantum Design Magnetic Property
Measurement System (MPMS-5).  The measurement of specific heat was
performed on on a Quantum Design Physical Property
Measurement System (PPMS-9).

The magnetic properties of LaCo$_{1-x}$Fe$_{x}$As was calculated using planewave basis pseudopotential method implemented in Quantum \emph{Espresso} package. The exchange-correlation interactions were modeled with Perdew, Burke and Enzerhoff flavor of
generalized gradient approximation\cite{lda}. To model the dilute substitutional iron doping
effect, a virtual crystal approximation (VCA) was employed to treat Fe/Co sites.

\section{\label{sec:level1}Results and Discussion}

\subsection{\textbf{Magnetic properties in LaCo$_{1-x}$Fe$_{x}$AsO}}

Fig. 1(a) shows the powder XRD patterns of the typical
LaCo$_{1-x}$Fe$_{x}$AsO samples and Fig. 1(b) shows the variations of
lattice parameters with respect to the Fe content ($x$). Where the
main diffraction peaks of those samples can be well indexed based on
a tetragonal cell of ZrCuSiAs-type structure, weak peaks exist due
to impurity phase CoAs. The content of impurity phase CoAs
estimated by Rietveld fitting is less than 5\%. It is worth noting
that CoAs has been reported to be non-magnetic from 4.2 to 300
K\cite{LaCoAsO}. The $a$-axis decreases slightly with increasing Fe content, and the $c$-axis increases accordingly, resulting in the
increase of the cell volume, since the ionic radius of tetrahedrally
coordinated Fe$^{2+}$ ions are larger than that of Co$^{2+}$. The
systematic increase in the $c$-axis indicates successful
substitution of Co by Fe. Similar variations of lattice constants
were also observed in the NdFe$_{1-x}$Co$_{x}$AsO in previous
reports\cite{NdFeCo}.

\begin{figure}[h]
\includegraphics[width=8cm]{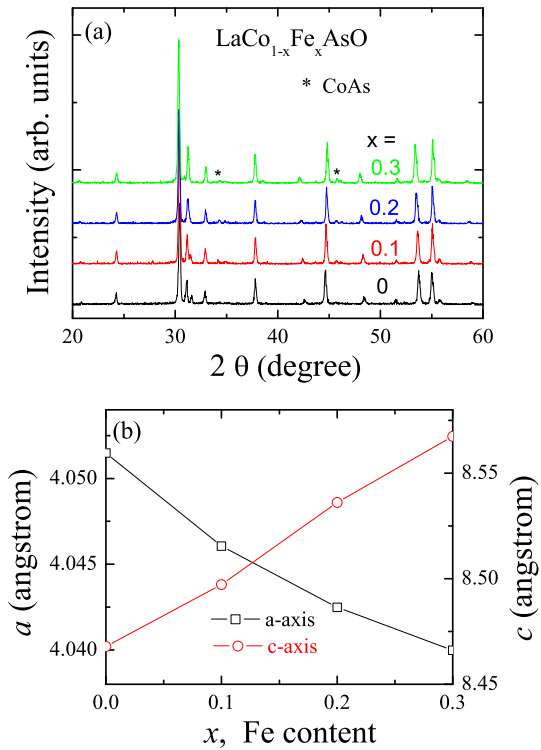}
\caption{\label{label}Structural characterization of
LaCo$_{1-x}$Fe$_{x}$AsO samples. (a) Powder X-ray diffraction
patterns of representative LaCo$_{1-x}$Fe$_{x}$AsO samples. The
asterisked peak positions designate the impurity phase of  CoAs.
(b) Lattice parameters as functions of Fe content.}
\end{figure}

Fig. 2 shows the temperature dependence of the electric resistivity ($\rho$) and magnetic
susceptibility ($\chi$) of LaCo$_{1-x}$Fe$_{x}$AsO samples. The inset shows magnetic susceptibility vs. temperature between 100 to 300 K. In Fig. 2(a), for LaCoAsO, the resistivity falls monotonically with decreasing temperature from 300 K, a resistivity hump can be clearly identified at about 63 K, which is related with FM transition temperature \emph{T$_{c}$}. As Fe content increases to 0.1, this hump shifts to about 35 K, and then for $x$ $=$ 0.2, the anomaly in resistive is not observed. Actually, the anomaly around \emph{T$_{c}$} becomes more obvious in the derivative of $\rho$ shown in Fig.2 (b), where \emph{T$_{c}$} decreases with increasing Fe content and shifts to below 2 K at $x$ $=$ 0.3. Meanwhile, the resistivity value gradually increases with the Fe doping levels, which can be attributed to the less itinerant nature
of Fe 3d electron than that of Co. The magnetic susceptibility  for the LaCo$_{1-x}$Fe$_{x}$AsO under \emph{H} $=$ 1 kOe in zero field cooled (ZFC) configuration was plotted in Fig. 2(c). For the parent LaCoAsO sample, the magnetic susceptibility
increases dramatically below \emph{T$_{c}$} of 63 K, suggesting that the Co sublattice forms
FM order in the CoAs layer. The similar magnetic behavior has been reported in the
literature\cite{LaCoAsO}. As Co is partially replaced by Fe, FM transition temperature (\emph{T$_{c}$}) is
sharply suppressed and shifts to lower temperature. For $x$ $=$ 0.3, the formation of long range FM
order cannot be identified in $\chi$(T) down to 2 K. Furthermore, the magnetic susceptibility value
drops to several orders of magnitude of LaCoAsO. On the other hand, it can be seen from inset that the magnetic susceptibility curve above 100 K exhibits the Curie-Weiss behaviors for LaCoAsO. As Fe content increases, the susceptibility gradually becomes less $T$-dependence and final remains constant for $x$ $=$ 0.3, indicating that Fe doping strongly reduces the moment of 3d-electrons in Co-based 1111 compounds.

\begin{figure}[h]
\includegraphics[width=8cm]{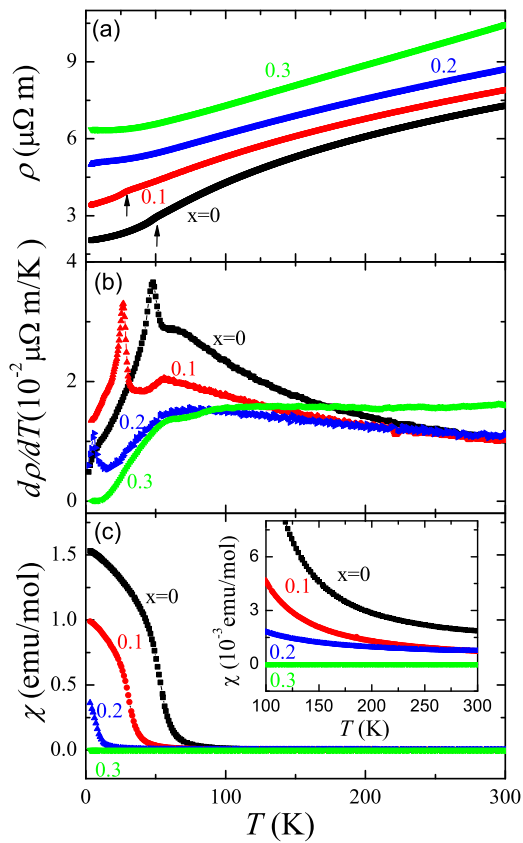}
\caption{\label{label}(a)Temperature dependence of resistivity
($\rho$) for the LaCo$_{1-x}$Fe$_{x}$AsO ($x$ = 0, 0.1, 0.2, 0.3) samples. (b) The derivative of resistivity near the ferromagnetic
phase transition. (c)Temperature dependence of magnetic susceptibility ($\chi$) under a magnetic field of 1000 Oe in the zero-field-cooled (ZFC) configuration.}
\end{figure}

Fig. 3 shows the M-H loop curves at several temperatures for LaCo$_{1-x}$Fe$_{x}$AsO. For $x$ $\leq$
0.2, the M-H curves are nearly linear above \emph{T$_{c}$}, indicating that those compounds are
paramagnetic at these temperatures. Below \emph{T$_{c}$}, these curves deviate from linearity
and become slightly S-shaped, suggesting the emergence of FM order. Further decreasing temperature
to 3 K, the molar magnetization sharply increases and then saturates with the increase of magnetic
field, and the small finite hysteresis can be distinguished (the data is not shown here). These results suggest that the ground state of these samples is FM. For $x$ $=$ 0.3, the M-H curve always shows the linear behaviors above 3 K, indicating that this sample remains paramagnetic. The magnetic moment\emph{M$_{s0}$} estimated by extrapolating the Ms-H curves to \emph{T} $=$ 0 K are 0.35 $\mu_{B}$ per Co for
LaCoAsO, which is very close to the value reported previously\cite{LaCoAsO}. With increasing Fe content, \emph{M$_{s0}$} quickly decreases, consistent with the fact that the FM transition temperature
\emph{T$_{c}$} shifts to lower temperature.

\begin{figure}[h]
\includegraphics[width=10cm]{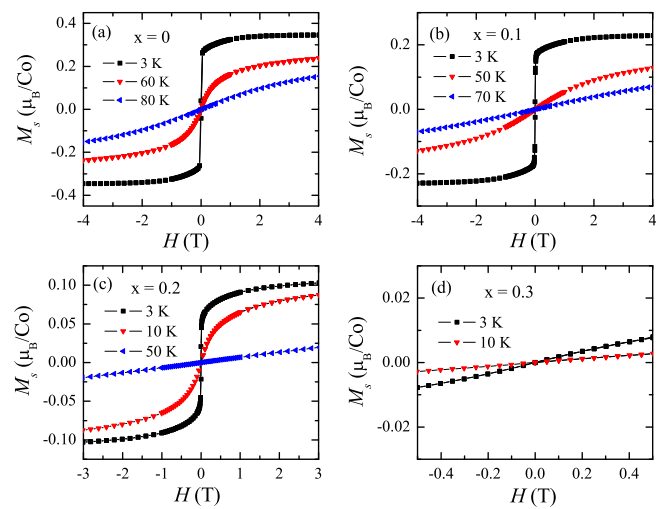}
\caption{\label{label}Field dependence of magnetization at various
temperatures  for the LaCo$_{1-x}$Fe$_{x}$AsO ($x$ = 0, 0.1, 0.2, 0.3)
samples.}
\end{figure}

Fig. 4 shows the LDA calculation results of LaCo$_{1-x}$Fe$_{x}$As system. The ground state
of the system were determined by comparing the total energy of three possible long range magnetic
orderings, i.e., ferromagnetic (FM), checkerboard antiferromagnetic (CB-AFM) and stripe-like
antiferromagnetic (SDW-AFM), as well as the non-magnetic (NM) configurations. As shown in
Fig. 4(a), the ground state is clearly FM at the cobalt side ($x$=0.0), where the
CB-AFM order cannot be stabilized over the whole range we considered ($0.0\leqslant x\leqslant
0.5$). As $x$ increases, the energy of FM configuration quickly rises, suggesting the iron doping
will suppress the formation of FM long-range order. At $x\sim$0.25, the SDW-AFM order takes over
and becomes the ground state. However, one should keep in mind that the disorder effect is not
fully taken care of in VCA, and a disordered dopant pattern is detrimental to the formation of AFM
long-range order. Furthermore, the magnetic coupling strength is suppressed with increasing $x$
from $x=$ 0.2 to 0.5, as suggested by increasing configuration energy of both FM and SDW-AFM orders.
With these considerations, we conclude that the actual ground state should be paramagnetic with
local magnetic fluctuations.

We then compare the LDA lattice structure variation in Fig. 4(b). As the experimental
measurements were performed at room temperature when the long range magnetic order was not yet
formed, we compare the NM lattice constants. It could be seen that between $x\in[0.0,0.3]$, the LDA
lattice constants show the same trend as the experimental results but with much larger variation.
Beyond $x =$0.3, the lattice constants show much less variation. For the FM phase of
LaCo$_{1-x}$Fe$_{x}$AsO, we also examine its magnetic moment per transition metal $m_{\mathrm{TM}}$
and the arsenic height $z_{\mathrm{As}}$ (Fig. 4(c)). At $x =$ 0.0, the LaOCoAs compound
has a small moment of 0.6 $\mu_B$/Co, indicating weak FM ground state with low Curie temperature.
As the doping level $x$ increases, the moment almost linearly decreases to $x$ $=$ 0.3 $\mu_B$/TM at
$x=$0.3 at the VCA level. As discussed above, the disorder effect will further suppress the moment.
Meanwhile, $z_{As}$ also increases almost linearly with respect to $x$ from $\sim$1.18\AA\ to
$\sim$1.24\AA. The variation of $z_{As}$ is consistent with the change of the super-exchange via
arsenic, which becomes less FM and more AFM with increasing $x$.

\begin{figure}[h]
\includegraphics[width=8cm]{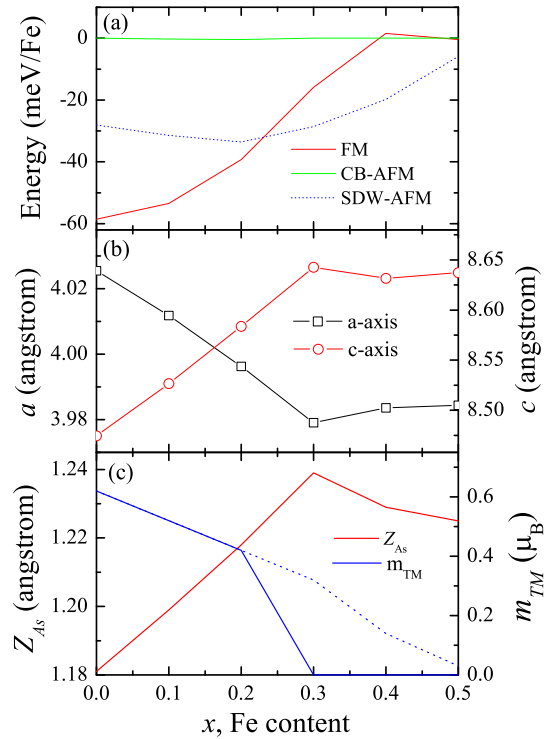}
\caption{\label{lda}LDA results of LaOCo$_{1-x}$Fe$_{x}$As properties. a) Magnetic configuration energy (per Fe atom) with respect to NM total energy. b) NM lattice constants variation with respect to $x$. c) Arsenic height $z_{\mathrm{As}}$ and transition metal magnetic moment $m_{\mathrm{TM}}$, the dashed blue line is the actual LDA results for FM phases, while the solid blue line suggests the actual scenario with phase transition taken into consideration.}
\end{figure}

\subsection{\textbf{Magnetic properties in SmCo$_{1-x}$Fe$_{x}$AsO}}

\begin{figure}[h]
\includegraphics[width=10cm]{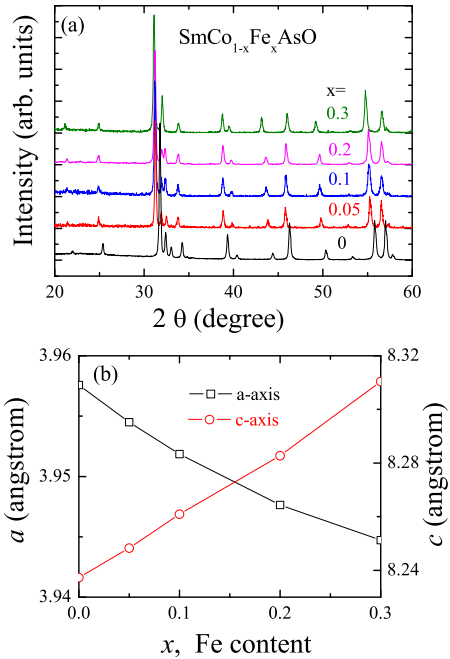}
\caption{\label{label}Structural characterization of
SmCo$_{1-x}$Fe$_{x}$AsO samples. (a) Powder X-ray diffraction
patterns of representative SmCo$_{1-x}$Fe$_{x}$AsO samples. (b)
Lattice parameters as functions of Fe content.}
\end{figure}
Fig. 5(a) shows the powder XRD patterns of SmCo$_{1-x}$Fe$_{x}$AsO samples and Fig. 5(b) shows the
variations of lattice parameters with Fe content ($x$). All those samples are single phase since no
extra peak is observed. Similar to the case of LaCo$_{1-x}$Fe$_{x}$AsO, Fe doping causes slight decrease in the
$a$-axis, while the $c$-axis monotonously increases.

\begin{figure}[h]
\includegraphics[width=8cm]{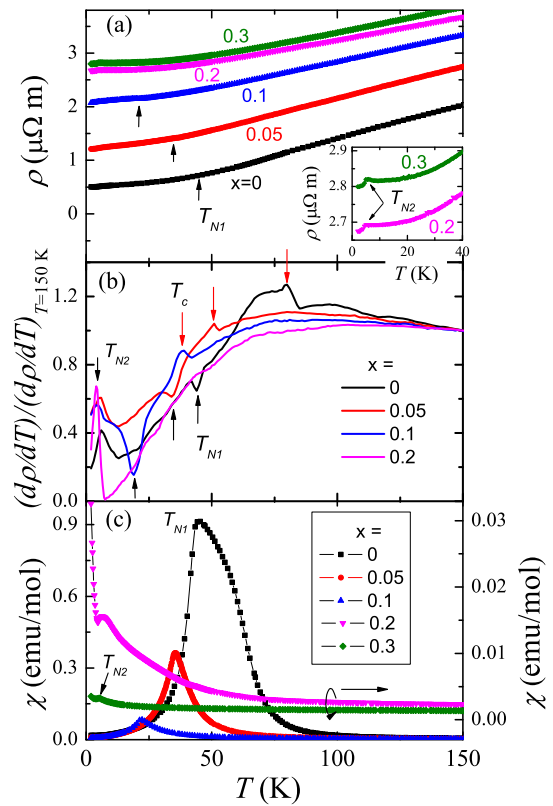}
\caption{\label{label}(a)Temperature dependence of resistivity
($\rho$) for the SmCo$_{1-x}$Fe$_{x}$AsO ($x$ = 0, 0.05, 0.1, 0.2, 0.3) samples. (b) The derivative of resistivity below 150 K. The data are normalized to $(d\rho/dT)_{T=150 K}$.  (c)Temperature dependence of magnetic susceptibility ($\chi$) under a magnetic field of 1000 Oe in the zero-field-cooled (ZFC) configuration. The inset shows the enlarged resistivity for x=0.2 and 0.3 at low temperature. }
\end{figure}

Fig. 6 shows the temperature dependence of the electric resistivity and magnetic susceptibility of
the SmCo$_{1-x}$Fe$_{x}$AsO samples. In Fig. 6(a), for the undoped parent compound
SmCoAsO\cite{LiCo}, the resistivity monotonically decreases with decreasing temperature from 300 K,
followed by a distinguishable kink around 45 K which can be associated with the FM-AFM transition
temperature (defined as \emph{T$_{N1}$)}. As Fe content increases to 0.1, this kink becomes more
pronounced and moves to lower temperatures, and no anomaly is observed below \emph{$T_{N1}$}. For $x$
$=$ 0.2, the resistivity anomaly related with \emph{T$_{N1}$} disappears, but another tiny kink can be
identified around 5.1 K, which can be attributed to the AFM transition (\emph{T$_{N2}$}) due to the
magnetic sublattice of Sm ions. As $x$ increases to 0.3, \emph{$T_{N2}$} in the resistivity becomes more
remarkable (shown in inset of Fig. 6(a)). In order to identify the magnetic ordering transition more clearly, the derivative of resistivity below 150 K is plotted in Fig.6 (b), which shows a maximum around \emph{T$_{c}$}, a minimum near \emph{$T_{N1}$}, and a peak at \emph{$T_{N2}$}. With increasing Fe content, the maximum around \emph{T$_{c}$} shifts to lower temperatures, and the minimum near \emph{$T_{N1}$} even becomes more pronounced for $x$ $=$ 0.1 and disappears as $x =$ 0.2. The magnetic susceptibility data is shown in Fig. 6(c), it can be seen that SmCoAsO shows a sharp peak around 45 K. This peak is ascribed to FM to AFM transition of the cobalt sublattice, which has been reported for \emph{Ln} $=$ Nd,
Sm, and Gd\cite{LnCoAsO}. As Fe substitutes Co, \emph{T$_{N1}$} is gradually suppressed and shifts to
lower temperatures. Meanwhile, the intensity of the peaks becomes weaker, consistent with the
resistivity data. At $x$ $=$ 0.2, it is noted that a tiny hump around 6.5 K is observed, which may
not be explained by the AFM transition \emph{T$_{N1}$} according to our M-H data (see fig.8). The detailed discussion
is beyond the scope of the current work and will be given in the future. When $x$ increases to 0.3,
the magnetization sharply drops and a tiny peak is detected at 5.5 K. Considering the specific heat
data in Fig. 7, this transition (\emph{T$_{N2}$}) is attributed to the AFM ordering of Sm$^{3+}$ sublattice\cite{AFM,Sm-AFM,Sm-AFM2}. Similar results can be found in the previous papers\cite{NdCoFe,Mcgurie,AFM}. For $x < 0.3$ samples, such low temperature peak in magnetization curves is not observable due to
the magnetic ordering of 3d--electrons.




\begin{figure}[h]
\includegraphics[width=8cm]{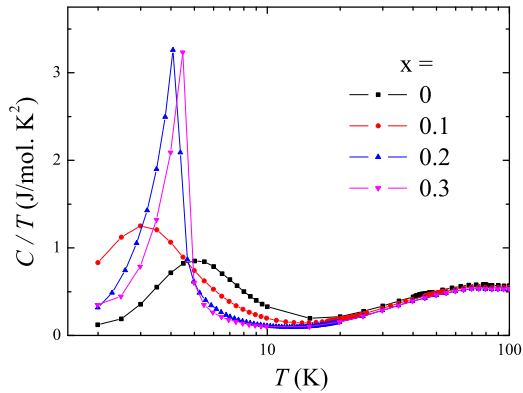}
\caption{\label{label} The specific heat of SmCo$_{1-x}$Fe$_{x}$AsO
($x$ = 0, 0.05, 0.1, 0.2, 0.3) samples under at zero field below 100
K.}
\end{figure}

In order to further study the magnetic phase transition, the zero field specific heat versus
temperature curves for those samples are summarized in Fig. 7. For all the samples, no anomaly in the
curves is found around \emph{T$_{c}$} associated with the FM transition, which is also the case in
NdCoAsO\cite{Mcgurie}. A small broad peak related to the AFM transition \emph{T$_{N1}$} can be observed at
45 K for SmCoAsO and then shifts to 22 K for $x$ $=$ 0.1 (The data is not shown here). However, it is
worth noticing that another clear broad peak from the AFM ordering of Sm$^{3+}$ sublattice can be
observed at 5.5 K and 3.5 K, respectively. As $x$ increases to 0.2, no extra peak is detected except
for the large anomaly near \emph{T$_{N2}$} of 5 K, which then shifts to 5.5 K for $x$ $=$ 0.3, consistent
with the magnetic susceptibility data shown in Fig. 6(c). Similar results are observed in the case of SmFeAsO\cite{AFM}, SmCoAsO\cite{SmCoAsO} and SmCoPO\cite{Sm-AFM2}.
However, the hump in SmCoAsO related to the Sm AFM ordering becomes sharper and higher with increasing Fe content, which suggests that the
increase of \emph{c} lattice weakens the coupling of 3d electron and 4f electron. Low temperature specific
heat measurements also confirms that the peak associated with \emph{$T_{N2}$} is robust as $x$ $\leq$ 0.3.

\begin{figure}[h]
\includegraphics[width=8cm]{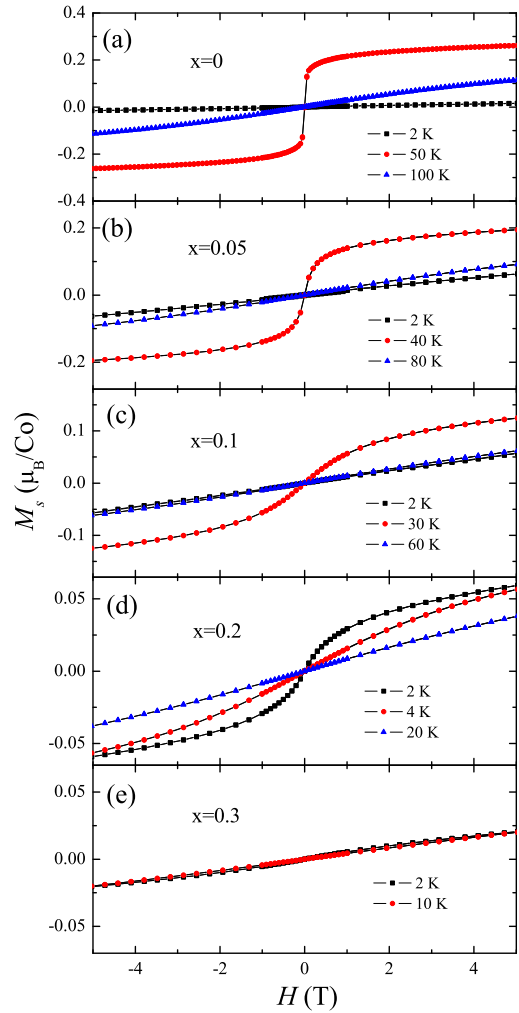}
\caption{\label{label}Magnetic field dependence of magnetization at several different temperature for the SmCo$_{1-x}$Fe$_{x}$AsO ($x$ = 0, 0.05 0.1, 0.2, 0.3) samples. }
\end{figure}

Fig. 8 shows the field dependence of the magnetization at various
temperatures for all the samples. As reported in previous
works\cite{SmCoAsO,LiCo}, SmCoAsO undergoes three magnetic phase
transitions at \emph{T$_{c}$}, \emph{T$_{N1}$}, and \emph{T$_{N2}$},
respectively. The magnetization sharply increases and saturates with
increasing field between \emph{T$_{c}$} and \emph{T$_{N1}$}. In
order to study the magnetic structure at different temperatures,
these data are collected at temperature where PM or FM state
dominates in Fig. 6(b), and 2 K. Obviously, for $x$ $\leq$ 0.1, the
$M-H$ curve is linear at 2 K and above \emph{T$_{c}$}, and the "S"
shape of FM behavior is only observed between \emph{T$_{c}$} and
\emph{T$_{N1}$}. As $x$ increases to 0.2, the linear feature is
observed only above \emph{T$_{c}$}, and the M-H curve displays FM
behavior below 2 K, implying that the AFM transition \emph{T$_{N1}$}
approaches to zero. This feature is different from the case of
NdCo$_{1-x}$Fe$_{x}$AsO\cite{NdFeCo}, where both FM and AFM order are not observed around 0.2. At $x$ $=$ 0.3, the M-H curves goes
back to the linear behavior at 2 K, indicating that the FM order
from the Co sublattice is completely destroyed. On the other hand,
with increasing Fe content, both the saturation moment and the
transition temperature (\emph{T$_{c}$}) gradually decrease, implying that the
3d--electron magnetism becomes weaker.

Based on above data, the magnetic phase diagram of
LnCo$_{1-x}$Fe$_{x}$AsO is established in Fig. 9. In
LaCo$_{1-x}$Fe$_{x}$AsO, the \emph{T$_{c}$} associated with FM
transition is sharply suppressed and disappears around $x$ $=$ 0.3. At
the same time, in the SmCo$_{1-x}$Fe$_{x}$AsO system, the \emph{T$_{N1}$}
from FM to AFM transition of Co sublattice is shifted to lower
temperatures with increasing Fe content and such transition has not
been observed at $x$ $=$ 0.2. Meanwhile the \emph{T$_{c}$}, the FM order
gradually decreases and is completely suppressed at $x$ $=$ 0.3,
similar to the case of LaCo$_{1-x}$Fe$_{x}$AsO. Here, we note that
for the Co-parent compounds \emph{T$_{c}$} increases slightly
when La is replaced by Sm. But a new FM-AFM transition of the
3d--electrons is induced in the later case. This manifests an
interesting interplay between the 3d--electrons and the local
Sm$^{3+}$ moments.

Thus, in the 1111--type Co-based LnCo$_{1-x}$Fe$_{x}$AsO systems, 4f
electrons of rare earth elements have an important effect on the
magnetic behavior of 3d electrons. Whereas, the antiferromagnetic
transition temperature of Sm moments\emph{T$_{N2}$} almost does not change
within the whole doping regime. It implies that the AFM ordering of
the Sm 4f--electrons is robust against Fe/Co substitution within the
CoAs layer. Therefore, the microscopic origin of the f-electron
AFM order of this system should be mainly due to the superexchange
interactions between the f-local moments\cite{DaiZhuSi}. These
superexchang interactions are bridged by two kinds of f--p
orbital hybridizations: one via Sm-O path and another the Sm-As
path, respectively\cite{DaiZhuSi}. Meanwhile, the RKKY interaction
mediated by the charge carries within the CoAs layer may not play a
crucial role for the f--electron magnetism because the RKKY
interaction would be explicitly dependent on the variations of
3d--electrons.  Because the radius of Sm$^{3+}$ ion is smaller than La$^{3+}$, one may speculate
that the FM-AFM transition is not related to the magnetic
4f--electrons, but rather due to the enhancement of three
dimensionality as the lattice parameter \emph{c} decreases. In order to
clarify this possibility, we have also performed LDA calculations on
LaCoAsO but using the lattice parameters of SmCoAsO. We do find that
the FM state of Co d--electrons is robust against the decreasing
lattice constant. Experimentally, Similar results were always observed in NdCoAsO under pressure\cite{pressure}. Therefore, we suggest that the FM-AFM transition at \emph{T$_{N1}$} of the 3d--electrons should be due to their coupling (polarization) to the 4f-moments along the $z$-direction.

\begin{figure}[h]
\includegraphics[width=8cm]{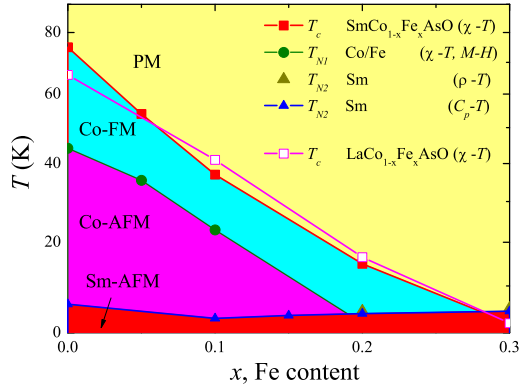}
\caption{\label{label}Magnetic phase diagram for the SmCo$_{1-x}$Fe$_{x}$AsO (solid) and LaCo$_{1-x}$Fe$_{x}$AsO (open) system, respectively. The transition temperatures were determined from the measurements of magnetic susceptibility, magnetization and specific heat. \emph{T$_{N2}$} is taken from resisitivity and specific heat data.}
\end{figure}

\section{\label{sec:level1}Conclusion}

In the 1111 Co-based \emph{Ln}Co$_{1-x}$Fe$_{x}$AsO (\emph{Ln}$=$La, Sm ) system, a series of \emph{Ln}Co$_{1-x}$Fe$_{x}$AsO samples
have been synthesized, and their transport and magnetic properties were investigated. A rich
magnetic phase diagram of the \emph{Ln}Co$_{1-x}$Fe$_{x}$AsO systems is then established. The FM order is
observed in both \emph{Ln}CoAsO (\emph{Ln} $=$ La,Sm) systems, and is completely destroyed with increasing Fe doping content to 0.3. Meanwhile, in SmCo$_{1-x}$Fe$_{x}$AsO, \emph{T$_{N1}$} is suppressed to below 2 K as $x$
$=$ 0.2, but the AFM order of rare earth element Sm ion survives in the whole doping regime $x$
$\leq$ 0.3. This also indicates that the disorder effect induced by Fe/Co doping is very weak.
Based on these results, it is concluded that in \emph{Ln}Co$_{1-x}$Fe$_{x}$AsO systems, while the magnetic
properties of the 4f electrons of rare earth elements are robust against the variations of 3d
electrons, they do play an significant role in the magnetic behaviors of the 3d electrons. These
materials therefore provide a prototypical testing ground for exploring the interplay between 4f
and 3d electrons in transition metal compounds.

\begin{acknowledgments}
Y.K. Li would like to thank J. H. Dai for discussions.  This work is
supported by the National Basic Research Program of China (Grant No.
2011CBA00103), NSFC (Grant No. 11174247, and 11104053),
and the National Science Foundation of Zhejiang Province(Grant No. Z6110033, and R12A040007).

\end{acknowledgments}

\end{document}